\documentclass[11pt,twoside]{article}
\usepackage{asp2006}
\usepackage{epsf}
\usepackage{graphicx}
\usepackage{psfig}
\usepackage{lscape} 

\begin{document}
\title{The Mass Loss Rates of sdB Stars}
\author{K. Unglaub}
\affil{Dr. Remeis-Sternwarte Bamberg, Sternwartstr. 7, D-96049 Bamberg,
Germany}
\begin{abstract}
According to previous investigations the effect of diffusion in the stellar 
atmospheres and envelopes of subdwarf B (sdB) stars with 
luminosities $10 \la L / L_{\odot} \la 100$ 
strongly depends on the presence of weak winds with mass loss rates  
$\dot M \la 10^{-12} M_{\odot}/ \rm yr$. These calculations with the 
mass loss rate as a free parameter have shown that it is hardly possible 
to reproduce the measured abundances of helium and metals simultaneously.
A possible reason is the decoupling of metals, which preferably absorb the 
photon momentum, from hydrogen and helium in the wind region.
In the present paper it will be investigated if ``chemically homogeneous" 
winds, as assumed in previous investigations, with mass loss rates 
$\dot M \leq 10^{-12} M_{\odot} / \rm yr$ can exist. 
From an observational point of view the existence of weak winds in sdB stars 
is unclear. Only in the most luminous ones possible wind signatures have 
been detected. Therefore it will be investigated if according to the theory 
of radiatively driven winds the existence of weak winds is plausible.
A stellar mass $M_{*} = 0.5 M_{\odot}$ is assumed.
  
The results for effective temperatures $T_{\rm eff} = 35000$, $30000$ and 
$25000 \rm K$, metallicities $0.1 \leq Z/Z_{\odot} \leq 1$ predict decreasing 
mass loss rates with increasing surface gravity. 
Dependent on the luminosity and metallicity the mass loss rates are between 
about $10^{-11} M_{\odot} / \rm yr$ and zero. If at all, chemically 
homogeneous winds can exist for the most luminous sdB stars only.
For the other ones selective winds are expected which should lead to 
additional changes of the surface composition.

In sdB stars, hot white dwarfs and HgMn stars (which are chemically peculiar 
main sequence stars) the measured metal abundances are tendencially lower 
than the ones predicted from diffusion calculations which assume an 
equilibrium between gravitational settling and radiative levitation.
Only for helium in almost all cases the measured abundances are larger 
than the predicted ones, but usually lower, below the solar value.
This may be an indication that the abundance anomalies of metals are 
preferably due to the selective winds, whereas the helium deficiencies 
are due to gravitational settling, which for still unknown reasons is 
less effective than expected in an undisturbed stellar atmosphere. 
\end{abstract}
\section{Introduction}
The abundance anomalies in subdwarf B (sdB) stars, white dwarfs and chemically peculiar main
sequence stars are believed to be at least partially due to the effect of diffusion 
in the stellar atmosphere and envelope. Several attempts have been made to explain 
the abundances with the effect of gravitational settling which may be counteracted by 
radiative levitation. As the radiative force on an element decreases with 
increasing abundance due to saturation effects, the surface composition can be predicted 
from the equilibrium condition between the inward gravitational force and the outward 
radiative force. However, an agreement between predicted and measured abundances 
requires the absence of disturbing processes like mass loss or convective mixing.
This may be the reason why in many cases the agreement is not satisfactory.
A comparison between predicted and measured metal abundances for sdB stars 
\citep[e.g.][]{t12_ber88,t12_ohl00,t12_chay06,t12_nat08}  
and hot white dwarfs \citep[e.g.][]{t12_son02,t12_son05,t12_good05}
shows that the measured abundances are tendencially 
lower than the predicted ones, although a few exceptions exist (e.g.\
 silicon in hot white dwarfs).

According to recent spectral analyses of sdB stars \citep[e.g.][]{t12_gei08,t12_gei08b,
t12_otol06,t12_edel06,t12_blan06}  especially the hotter ones with 
$T_{\rm eff} > 30000 \rm K$ in many cases show strong deficiencies  
of light metals like Al, Mg, O and Si by more than a factor of $100$ in 
comparison to the solar abundances, whereas enrichments of elements heavier 
than iron by a factor of $100$ are not unusual. Qualitatively, these abundance 
patterns show some similarity to those found in HgMn stars, which are a subgroup 
of the chemically peculiar main sequence stars reviewed by \citet{t12_smith96}. 
The HgMn stars with $10000 \rm K \la T_{\rm eff} \leq 16000 \rm K$, $\log g \approx 4.0$ 
are characterized by low rotational velocities and weak or non-detectable magnetic 
fields. Some light metals (e.g.\ Al, N) tend to be deficient, whereas  
heavy metals (e.g.\ Hg, Mn, Pt, Sr, Ga) may be enriched up to several orders of magnitude. 
Some of the 
most recent spectral analyses are from \citet{t12_zav07} and 
\citet{t12_adel06}. The measured abundances of iron group elements 
\citep{t12_sea96,t12_jom99} as well as nitrogen \citep{t12_roby99} tend to be lower than 
predicted from equilibrium diffusion calculations. Only for mercury abundances larger than 
predicted have been detected \citep{t12_prof99}.

So in hot white dwarfs, sdB and HgMn stars a common tendency seems to be present, according 
to which the metal abundances are lower than expected from the equilibrium condition between 
gravitational settling and radiative levitation. In addition there is a large scatter of 
abundances from star to star. Even for stars with similar stellar parameters the abundances 
may be different. This points to some time-dependent process and not to an equilibrium 
state. In main sequence stars the chemically peculiar phenomenon is restricted to stars 
with low rotational velocities. In addition the presence of magnetic fields may be of
importance, because magnetic fields may change the radiative acceleration or suppress 
convection \citep{t12_tur03}. White dwarfs and sdB stars, however, are always more or less 
chemically peculiar. Up to now no correlation between abundance anomalies and magnetic 
field strengths has been found \citep{t12_otol05}. In hot DA white dwarfs no outer convection
zones should exist, because hydrogen is preferably ionized and helium is strongly 
deficient. In sdB stars a thin superficial convection zone with a mass depth of the 
order $10^{-12} M_{*}$ may be present only for helium abundances 
$\rm He / \rm H \geq 0.01$ by number \citep{t12_gro85}. So it seems to be unlikely 
that magnetic fields are of decisive importance for the explanation of the surface 
compositions.

In hot DAO white dwarfs helium is detectable, but in many cases it is deficient in comparison 
to the solar value \citep[e.g.][]{t12_napi99}. The same is true for the majority of sdB's 
\citep[e.g.][]{t12_edel03,t12_lisk05} 
and the helium deficient main sequence stars (e.g.\ in HgMn stars helium usually is
deficient). In contrast to the metal abundances, however, the abundances of helium are 
always larger than predicted from equilibrium diffusion calculations 
\citep{t12_ven88,t12_mic89,t12_mic79}. These results may be explained with the effect of gravitational
settling which, however, somehow must be disturbed. Possible reasons 
for this disturbance of the equilibrium may be the presence of turbulence \citep{t12_vauc78} 
or mass loss. \citet{t12_fon97} and \citet{t12_ub98} predicted helium abundances as a function 
of time in the presence of weak winds. The results have shown that for mass loss rates of the
order $10^{-13} M_{\odot} / \rm yr$ helium sinks much more slowly than in the case of an 
undisturbed stellar atmosphere. Within the lifetimes of sdB stars near the extended 
horizontal branch ($\approx 10^{8} \rm yr$) the helium abundance would gradually decrease 
from the solar value to $\rm He / \rm H \approx 10^{-4}$ by number. This could explain why 
the helium abundances usually are in this range.

If this scenario with weak winds were the correct explanation for both
the helium and the metal abundance anomalies, then it should be
possible to find a mass loss rate for which all abundances can be
explained simultaneously. The calculations of \citet{t12_ub01} for the
elements H, He, C, N and O have shown that this is hardly
possible. According to these calculations for solar initial
composition helium should always be more deficient than the metals. No
mass loss rate exists which leads to deficiencies of C and O by more
than a factor of $100$, whereas helium is deficient by a factor of ten
only. This, however, is not an unusual composition in sdB stars
\citep[e.g.][]{t12_heb00}. Moreover in the presence of winds with mass
loss rates of the order $10^{-13} M_{\odot} / \rm yr$ which are
required to explain the helium abundances, the proposed pulsation
mechanism of some sdB stars \citep{t12_char97,t12_fon03} would become
questionable. As mass loss tends to level out concentration gradients,
the reservoir of iron (or other iron group elements) in the stellar
envelope needed to explain the pulsations should be destroyed in time
scales which are much shorter than the lifetimes of sdB
stars. According to \citet{t12_chay04} and \citet{t12_fon06} for a mass loss
rate of $\dot M = 6\times 10^{-15} M_{\odot} / \rm yr$ the reservoir
would be destroyed on about $10^{7} \rm yr$. For $\dot M = 10^{-13}
M_{\odot} / \rm yr$ the matter in mass depths $\la 10^{-7} M_{*}$,
where the reservoir is expected, would be blown away within one
million years only.
 
Probably the most crucial assumption in these diffusion calculations
with mass loss has been that the winds are ``chemically
homogeneous". If $\dot M_{l}$ is the mass loss rate of an element $l$ and
$\zeta_{l}$ its mass fraction in the photosphere, then this assumption
states that $\dot M_{l} = \zeta_{l} \dot M$, where $\dot M$ is the
total mass loss rate.  Such a chemically homogeneous wind prevents (if
$\dot M \ga 10^{-11} M_{\odot} / \rm yr$) or retards (if $\dot M <
10^{-11} M_{\odot} / \rm yr$) gravitational settling. However, it does
not directly change the surface composition. The opposite case would
be a ``selective" wind in which the mass loss rates of the individual
elements are essentially independent of each other. A selective wind
should lead to additional changes of the surface composition, which
have not yet been taken into account in the calculations.

In radiatively driven winds of hot stars the photon momentum is absorbed preferably by 
the metals \citep[see e.g.][]{t12_abb82,t12_vink01}, whereas the contribution of 
hydrogen and helium is small. Thus the metals are accelerated and move outwards. 
If the flow of metals is sufficiently large, then due to Coulomb collisions with the 
metals hydrogen and helium are accelerated as well. For this purpose, in dense winds 
a small velocity difference between the metals and hydrogen and helium is sufficient.  
Then it should be a good approximation that all elements have the same velocity and that 
the wind is chemically homogeneous. If, however, the flow of metals is small, then it 
may happen that the coupling of the various constituents due to collisions is not 
sufficiently effective and that hydrogen and helium are left behind. This scenario 
may lead to pure metallic winds such as investigated by \citet{t12_bab95} for main sequence 
A stars.

Mass loss has been detected in subdwarf O stars (sdO) 
which are more luminous than sdB's 
\citep{t12_ham81,t12_rau93}. 
Up to now there is no observational proof for the existence of winds 
in sdB stars. From a quantitative analysis of $\rm H \alpha$ line 
profiles of 40 sdB stars \citep{t12_max01}, a comparison of 
synthetic NLTE $\rm H \alpha$
line profiles from static model atmospheres with the observations revealed 
perfect matches for almost all stars. Only in the four most luminous sdB's 
anomalous $\rm H \alpha$ lines with a small emission at the line center have 
been detected, which possibly are signatures of weak winds 
\citep{t12_heb03}.
For the case $T_{\rm eff} = 36000 \rm K$, $\log g = 5.5$ and $\log L/L_{\odot} = 
1.51$, from a spectral synthesis of $\rm H \alpha$ with his wind code 
\citet{t12_vink04} found a similar behaviour of $\rm H \alpha$ if the existence of 
a weak wind with $\dot M \approx 10^{-11} \rm M_{\odot} / \rm yr$ is assumed.

In Sect.~2 it will be investigated for 
a stellar mass $M_{*} = 0.5 M_{\odot}$  if winds 
with mass loss rates $\dot M \la 10^{-12} M_{\odot} / \rm yr$ can be 
chemically homogeneous. The arguments are similar as in the 
investigations for more luminous stars e.g.\ from \citet{t12_ow2002}, 
\citet{t12_spr92} and \citet{t12_krt03}. All metals are lumped together into one 
mean metal which is accelerated due to the absorption of photon momentum.
This is some simplification, in \citet{t12_krt06} the individual elements are 
considered separately. 
Hydrogen, helium and the free electrons are denoted as ``passive plasma" which can 
only be accelerated due to collisions with the outflowing metals.
From the calculations e.g.\ of \citet{t12_abb82} and \citet{t12_vink01} as well as 
from own calculations as described in Sect.~3 there is no indication that 
the radiative force on hydrogen and helium is not negligible. The mean 
radiative acceleration on the metals exceeds the one on hydrogen and helium 
by at least a factor of $100$.

In Sect.~3 the results of mass loss calculations for sdB stars according to 
the original theory of radiatively driven winds of \citet{t12_ca1975} are 
presented and are compared with the predictions from the mass loss recipe 
of \citet{t12_vink02}. For these wind models, which are obtained from a one 
component description of the wind, it is again checked if the metals may 
be coupled to hydrogen and helium. In Sect.~4 the consequences of the results 
for the surface composition of sdB stars are discussed.

\section{Do Chemically Homogeneous Weak Winds Exist?}
In the following the hydrogen and helium, for which the radiative acceleration is 
assumed to be zero, will be denoted as ``element" 1, whereas the mean metal is denoted 
as ``element" 2. As in the supersonic region the gradient of the gas 
pressure can be neglected, the momentum equation for ``element" 1 (H+He) 
can be written as:
\begin{equation}
g_{\rm coll} = \frac {G M_{*}}{r^{2}} + v_{1} \frac {dv_{1}}{dr}
\end{equation}
$g_{\rm coll}$ is the collisional acceleration on hydrogen and helium due to Coulomb 
interactions with the metals and $v_{1}$ is the mean velocity of 
hydrogen and helium. The existence of a supersonic wind 
with increasing velocity in outward direction requires that the 
acceleration term $v_{1} \frac {dv_{1}}{dr}$ is larger than zero. 
Thus this equation can only be valid if: 
\begin{equation}
g_{\rm coll} > \frac {G M_{*}}{r^{2}}
\end{equation}
where $G$ is the gravitational constant and $r$ the radius.
Hydrogen and helium cannot be accelerated if this condition is violated. 
As derived by \citet{t12_bur69}, $g_{\rm coll}$ can be written as:
\begin{equation}
g_{\rm coll} = \frac {\rho_{2}}{m_{1} m_{2}} \frac {4 \pi Z_{1}{^2} Z_{2}{^2} e^{4}} 
{k T} \left ( \ln \Lambda \right ) G_{\left ( x \right )}
\end{equation}
$m_{1}$ and $m_{2}$ are the mean masses of ``elements" 1 and 2, 
$\rho_{2}$ is the mass density of the mean metal, $k$ the Boltzmann constant, 
$T$ the temperature and $e$ is the electron charge (cgs system). 
$Z_{1}$ is the mean charge 
of hydrogen and helium, $Z_{2}$ is the charge of the mean metal.
$\ln \Lambda$ is the Coulomb logarithm which according to 
\citet{t12_bur69} is:
\begin{equation}
\ln \Lambda = - 1/2 
+ \ln \left (  \frac {3kT R_{\rm D}}{Z_{1} Z_{2} e^{2}} \right )\,.
\end{equation}
$R_{\rm D}$ is the Debye radius:
\begin{equation}
R_{\rm D} = \left ( \frac {k T}{4 \pi e^{2} \left ( n_{1} Z_{1}^{2}
+ n_{2} Z_{2}^{2} + n_{\rm e} \right )} \right )^{0.5}
\end{equation}
where $n_{1}$ is the number density of hydrogen and helium, 
$n_{2}$ and $n_{\rm e}$ are the number density of the mean metal and the 
free electrons, respectively. 

$G_{\left ( x \right )}$ in Eq.~3 is the Chandrasekhar function defined as, 
e.g., in \citet{t12_krt03}. It depends on the quantity $x$, which approximately is 
the velocity difference of the mean metal and the passive plasma in units of 
the thermal velocity of hydrogen. If $v_{2}$ is the velocity of the mean metal 
and $v_{1}$ the mean velocity of the passive plasma, then the latter can be 
accelerated only if $v_{2} - v_{1} > 0$. For small velocity differences, 
$G_{\left ( x \right )}$ and thus the collisional acceleration approximately 
is proportional to $v_{2} - v_{1}$. If, however, the velocity difference is near 
the thermal velocity of hydrogen, $G_{\left ( x \right )}$ reaches a maximum value 
and decreases to larger velocity differences. For the discussion in 
the present section only this maximum value is important:
\begin{equation}
G_{\rm max} = 0.214\,.
\end{equation}

From Eq.~3 it can be seen that the collisional acceleration on hydrogen and 
helium is proportional to $T^{-1}$. In the present paper $T = T_{\rm eff}$
is assumed for the wind region. If the temperature were larger, e.g.\ due to 
frictional heating, this would reduce $g_{\rm coll}$ and thus favour the decoupling 
of the metals. This effect could (partially) be compensated, if the heating 
leads to a higher mean charge of the metals, because $g_{\rm coll} \sim Z_{2}^{2}$.
Thus in this case more detailed calculations would be necessary.

The mass density $\rho_{2}$ of the mean metal in Eq.~3 can be substituted 
from the equation of continuity for the metal:
\begin{equation}
\rho_{2} = \frac {\dot M_{2}}{4 \pi r^{2} v_{2}}
\end{equation}
where $\dot M_{2}$ is the mass loss rate of the metals and $v_{2}$ their mean 
velocity. With this substitution condition Eq.~2 can 
be transformed into a condition 
for $\dot M_{2}$:
\begin{equation}
\dot M_{2} > \frac {m_{1} m_{2} k T}{Z_{1}^{2} Z_{2}^{2} e^{4} \ln \Lambda\,
 G_{\rm max}}G M_{*} v_{2}\,.
\end{equation}
This condition depends on the mean velocity of the metals. It is most 
restrictive in the part of the wind where $v_{2}$ is maximal.
In accelerating winds this maximal velocity is the terminal 
velocity $v_{\infty}$.
According to theoretical calculations and observations of hot star 
winds \citep[see e.g.][]{t12_la1999}, radiatively driven winds 
accelerate to terminal velocities which are at least of the order of 
the surface escape velocity $v_{\rm esc}$. Thus we require that this 
condition must still be fulfilled for
\begin{equation}  
v_{2} = v_{\rm esc} = \sqrt {\frac {2 G M_{*}}{R_{*}}} 
= \sqrt {2} \left ( G M_{*} g \right )^{\frac {1}{4}}\,.
\end{equation}
Here it has been assumed that the radiative force due to electron 
scattering is negligible which is justified for subluminous stars. 
The right equation is obtained from $R_{*} = \sqrt {G M_{*} g^{-1}}$.
With this expression for $v_{2}$ condition (8) can be written as:
\begin{equation}
\dot M_{2} > \frac {m_{1} m_{2} k T}{Z_{1}^{2} Z_{2}^{2} e^{4} \ln 
    \Lambda\, G_{\rm max}}
\sqrt {2} \left ( G M_{*} \right )^{\frac {5}{4}}  g^{\frac {1}{4}}\,.
\end{equation}
Now we assume a mean mass for hydrogen and helium $m_{1} = m_{\rm p}$ (where 
$m_{\rm p}$ is the proton mass) and $m_{2} = 15 m_{\rm p}$ for the mean metal.  
The mean charges are assumed to be $Z_{1} = 1$ and $Z_{2} = 3$, respectively. 
According to own calculations $\ln \Lambda$ may vary between about $6.0$ at the 
wind base and $18.0$ in the outermost regions. As an upper limit and 
similar to \citet{t12_ow2002} we assume $\ln \Lambda = 20$. 
With a typical stellar mass for sdB stars $M_{*} = 0.5 M_{\odot}$ 
and for $T = T_{\rm eff}$ it follows:
\begin{equation} 
\dot M_{2} > 1.2 \times 10^{-20} T_{\rm eff} g^{\frac {1}{4}}\,.
\end{equation}
The mass loss rate $\dot M_{2}$ of the metals is given
in $M_{\odot} / \rm yr$, $T_{\rm eff}$ in K and the surface 
gravity $g$ in $\rm cm \, \rm s^{-2}$.
For typical stellar parameters of sdB stars, e.g.\ 
$T_{\rm eff} = 30000 \rm K$, $\log g = 5.7$, it follows
\begin{equation} 
\dot M_{2} \ga 10^{-14} M_{\odot} / \rm yr \,.
\end{equation}
This means that the mass 
loss rate of the metals alone must exceed this value. Otherwise hydrogen 
and helium will decelerate and cannot be expelled from the star, if not 
the decoupling occurs at such a large radius at which the velocity is already 
larger than the local escape velocity.

If the wind is chemically  homogeneous, this implies  
$\dot M_{2} = \zeta_{2} \dot M$. If the mass fraction of the metals is of 
similar order of magnitude as the solar one ($\zeta_{2} \approx 0.01$), then 
the condition $\dot M_{2} \ga 10^{-14} M_{\odot} / \rm yr$ implies for the 
total mass loss rate $\dot M \ga 10^{-12} M_{\odot} / \rm yr$. Thus chemically 
homogeneous winds with lower mass loss rates cannot exist, if the metallicity 
is not larger than solar. A similar result can be obtained from 
eq.~22 
of \citet{t12_ow2002}. In this paper a maximum velocity $v_{\rm max}$ is derived 
up to which the constituents may be coupled. If the wind accelerates at least 
to the surface escape velocity, then a necessary condition for a the existence 
of a coupled wind is $v_{\rm max} > v_{\rm esc}$.
\section{Mass Loss Rates from a One Component Description of the Wind}
In this section it will be discussed how large the mass loss rates are in 
sdB stars according to the theory of radiatively driven winds.
These calculations will be described in more detail in a forthcoming paper 
\citep{t12_ku08}.
The momentum equation for an isothermal wind is 
solved without the usual parametrization of the line force multiplier parameters, 
which becomes questionable in weak winds \citep{t12_ku2002}. As later improvements of 
the CAK theory  (finite disk correction and changes of ionization 
in the wind) are 
neglected, the method of solution is straightforward. The omission of the 
finite disk correction leads to an overestimate of $\dot M$ by a factor of the 
order two to three \citep{t12_la1999}. From these calculations, in which 
multicomponent 
effect are neglected, the total mass loss rate $\dot M$ and the velocity law 
$v_{\left ( r \right )}$ are derived. From this solution the acceleration term 
$v \frac {dv}{dr}$ is known. With the assumption that metals are trace elements 
($v_{1} \approx v$) it can be checked if metals may be coupled to hydrogen and 
helium from a criterion similar to the one derived by \citet{t12_ow2002}. This criterion
requires that hydrogen and helium must accelerate. Thus it will predict decoupling 
at larger mass loss rates  than the less restrictive criterion Eq.~2, 
which only 
requires that the passive plasma does not decelerate.  
\begin{figure}[htp!]
\centering
\includegraphics[width=8.5cm,bb=105 70 625 450,clip=true]{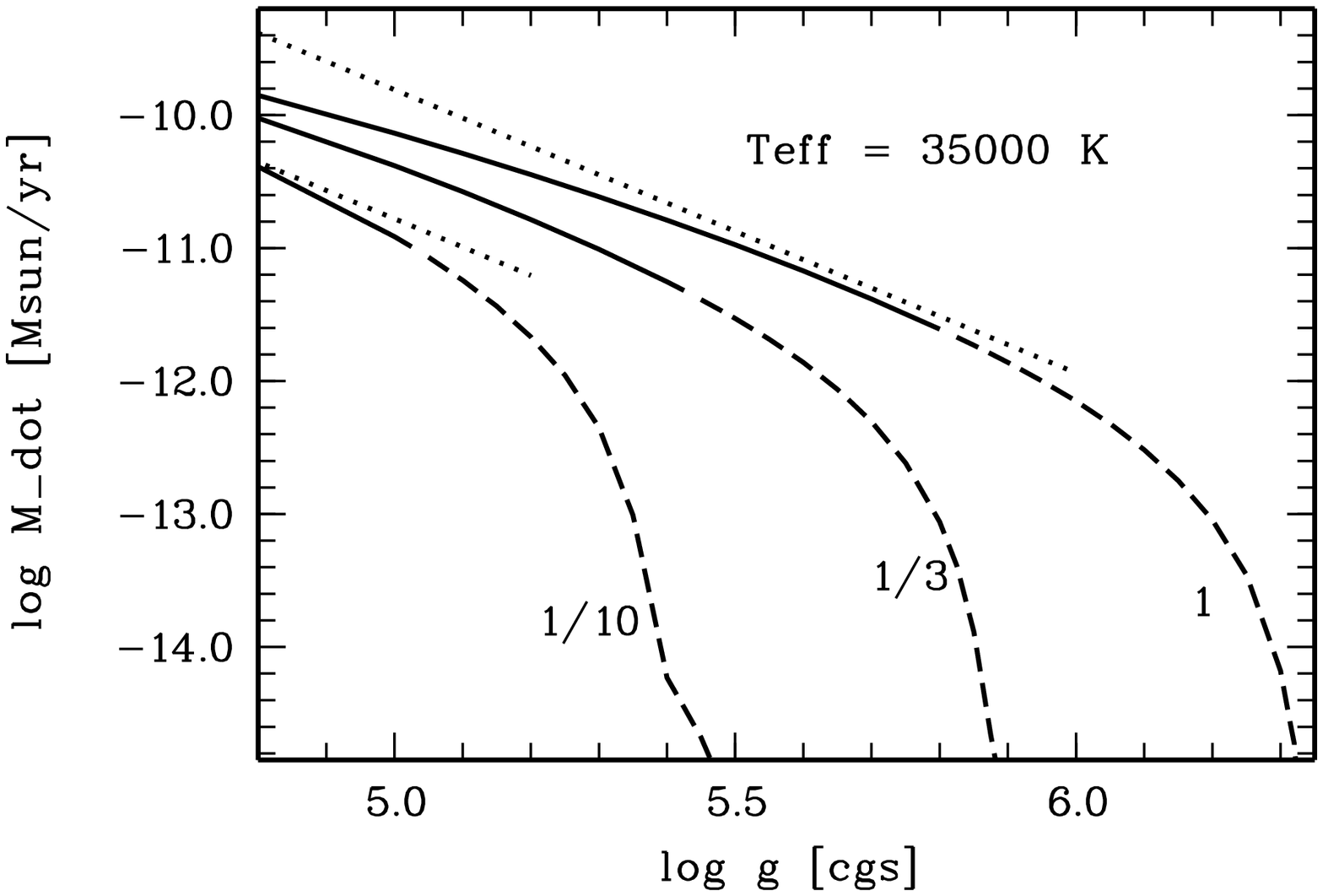}
\includegraphics[width=8.5cm,bb=105 70 625 450,clip=true]{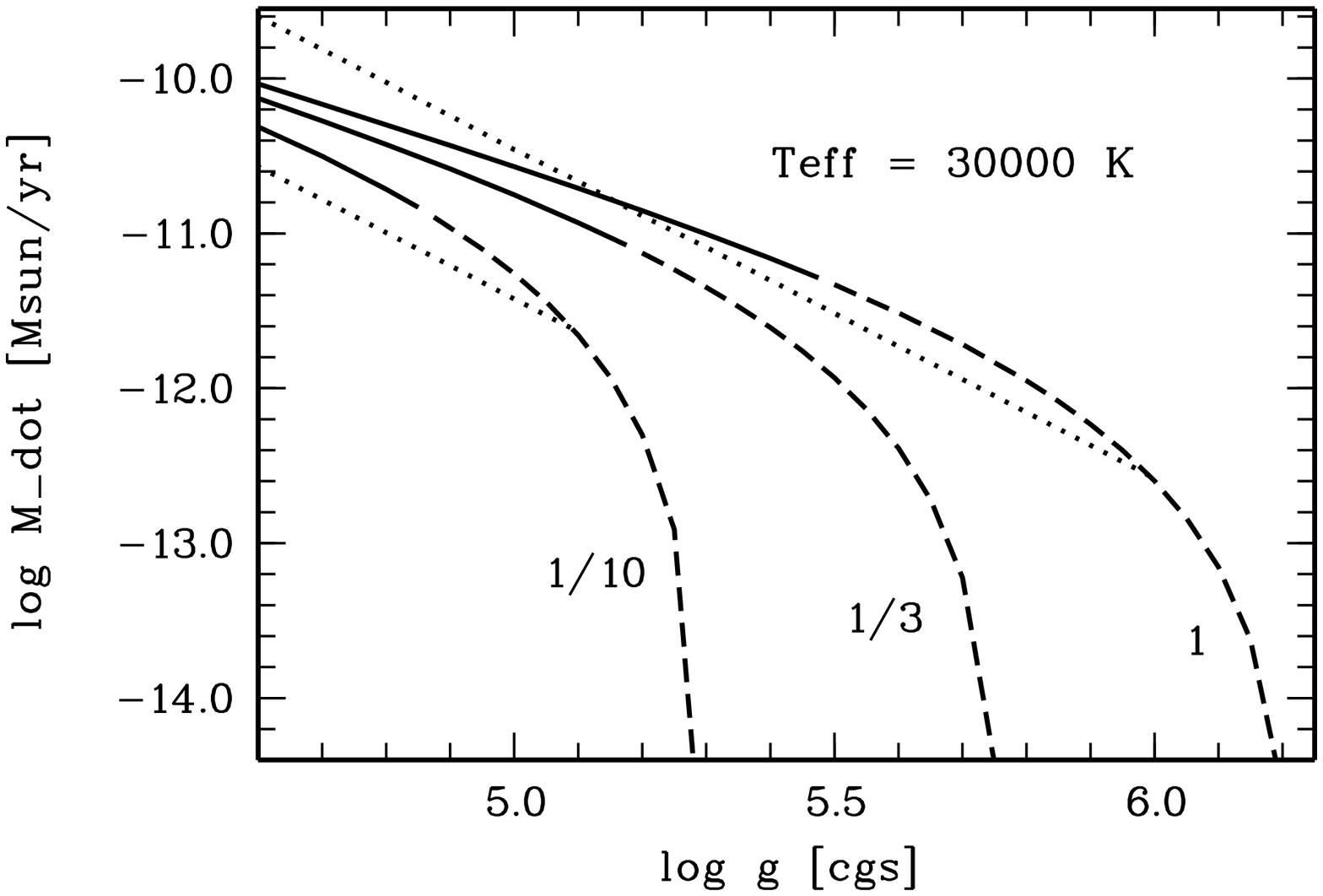}
\includegraphics[width=8.5cm,bb=105 70 625 450,clip=true]{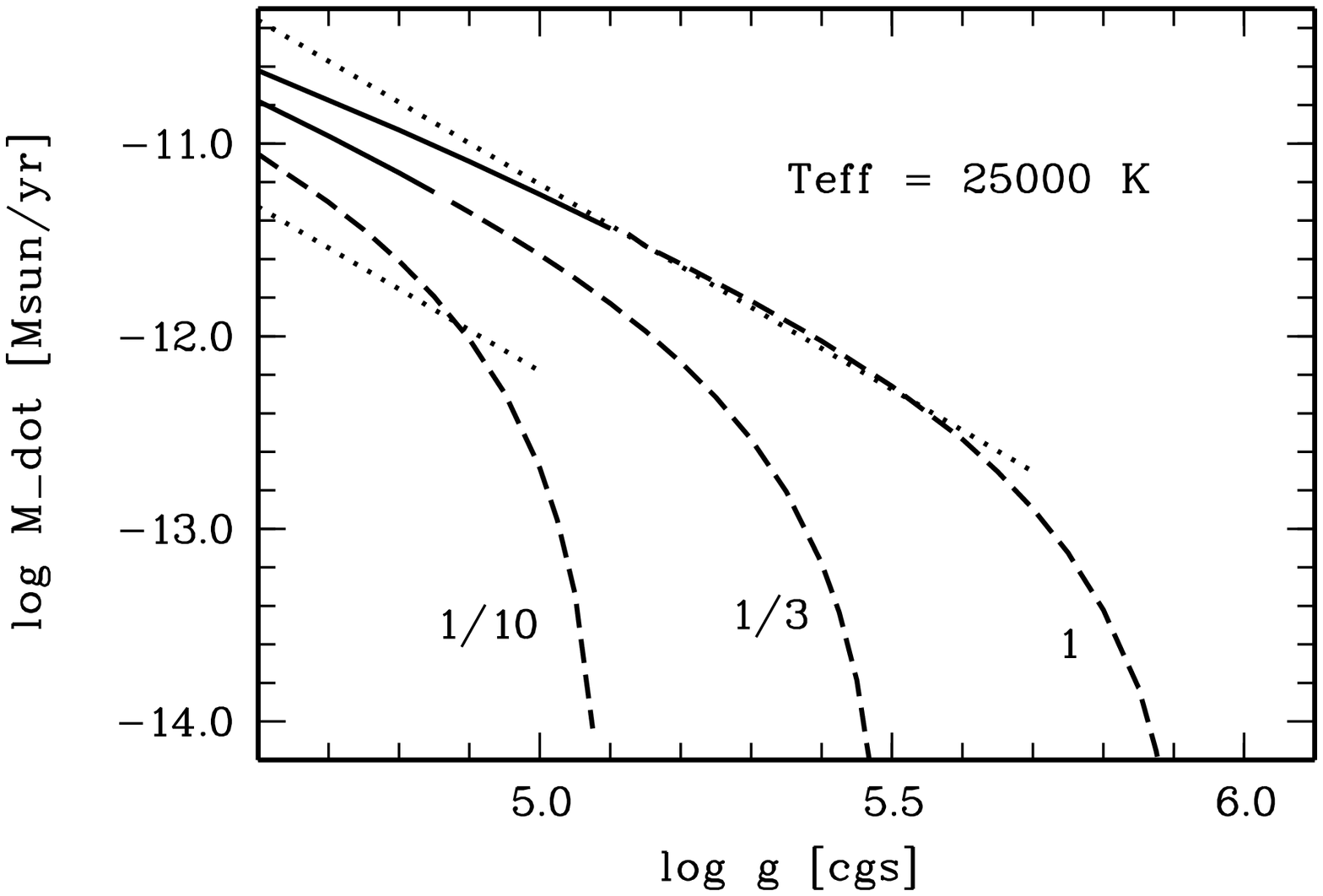}
\caption{Predicted mass loss rates as a function of surface gravity for 
$T_{\rm eff} = 35000$, $30000$, $25000 \rm K$ and $Z/Z_{\odot} = 1$, 
$1/3$, $0.1$ with $M_{*} = 0.5 M_{\odot}$. Dashed lines indicate decoupling 
of metals in the corresponding wind model. The upper and lower dotted lines 
in each figure represent the results of Vink \& Cassisi \citep{t12_vink02}.}
\end{figure}

In Fig.~1 the predicted mass loss rates are shown as a function of
surface gravity for $T_{\rm eff} = 35000$, $30000$ and $25000 \, \rm
K$ and several metallicities: $Z / Z_{\odot} = 1$, $1/3$ and
$1/10$. For $Z / Z_{\odot} = 1$ and $1/10$ the results are compared to
the ones of \citet{t12_vink02}, which are represented by dotted lines.  It
can be seen that especially for $Z/Z_{\odot} = 1$ the agreement is
very well in many cases. This, however, is a consequence of the
assumptions.  As in the present calculations the original version of
the CAK theory is used, for the terminal velocity it follows
$v_{\infty} \approx v_{\rm esc}$ and the solution of the momentum
equation has the form 
\begin{displaymath}
v_{\left ( r \right )} = v_{\infty} \left ( 1 -
\frac {R_{*}}{r} \right )^{0.5}\,.
\end{displaymath}
\citet{t12_vink02} assumed a similar
velocity law with $v_{\infty}$ as a free parameter.  Their results
shown in Fig.~1 are for $v_{\infty} = v_{\rm esc}$. With this
$v_{\left ( r \right )}$ they obtain the mass loss rate from the
requirement of a global momentum conservation.  As the function
$v_{\left ( r \right )}$ approximately is the same in both
calculations, the agreement of the mass loss rates should be
expected. Nevertheless it may appear surprising, because Vink \&
Cassisi's calculation of the radiative acceleration is clearly more
sophisticated than the present one. They took into account about
$10^{5}$ lines of the elements H--Zn with NLTE occupation numbers for
the most important elements, whereas in the present calculations only
the elements H, He, C, N and O have been taken into account with a
line list similar to the one used in the diffusion calculations of
\citet{t12_vauc79}. It consists of about $150$ lines only and the
occupation numbers have been obtained with the assumption of LTE,
which is not a good approximation at low densities in the wind region.

The main reason why in spite of these very different assumptions the
predicted mass loss rates are similar is, that in weak winds the major
part of the radiative force is due to a few strong lines only
\citep{t12_abb82,t12_puls00}. E.g.\ for $T_{\rm eff} = 30000 \rm K$, $\log g =
5.5$, $Z/ Z_{\odot} = 1$ the predicted mass loss rate according to the
present calculations is $\dot M = 4.7 \times 10^{-12} M_{\odot} / \rm
yr$. In this example, about 70 \% of the radiative acceleration are
due to three carbon lines only: these are the lines C\,III 
$\lambda 977$\,\AA\ and $\lambda 1176$\,\AA\ and 
C\,IV $\lambda 1549$\,\AA\ (doublets or
multiplets are considered as one line).  If only these three lines
were taken into account, the mass loss rate would be lower by a factor
of two only. This means that the result essentially depends on the
abundance of carbon. With the present assumptions 58 \% of all carbon
is C\,III and 42 \% is C\,IV. If the C\,III/C\,IV ionization equilibrium is
changed, then the mass loss rates would still be of similar order of
magnitude. With the assumption that all carbon is in the ground state
of C\,III so that from the considered three lines only $\lambda 977$\,\AA\
contributes, the predicted mass loss rate would be lower by a factor
of two. If all carbon were in the ground state of C\,IV, so that only
$\lambda 1549$\,\AA\ contributes, $\dot M$ would decrease by a factor of
three.  So independently how this ionization equilibrium is, at least
the order of magnitude of the mass loss rates will always be similar.

The fact that the wind is driven by a few strong lines may have important 
consequences, however. In the CAK theory it is assumed that the stellar flux
which can be absorbed by the matter in the wind is independent of the velocity. 
This is justified in dense winds which are driven by a large number preferably 
of weak lines. In the present calculations the emergent flux of models as described  
e.g.\ in \citet{t12_ub01} has been used with a temperature structure according 
to the diffusion approximation for the flux. This flux is in good agreement with 
the continuum fluxes of model atmospheres of sdB stars (Heber, priv. comm.).
Near the center of strong lines, however, the flux may be lower by a factor of the order 
ten. Thus in the inner parts of the wind, where the velocity and thus 
the Doppler shift is small, the present calcuations clearly overestimate the stellar 
flux and thus the radiative acceleration. According to \citet{t12_bab96} this effect 
of ``line shadowing" leads to lower mass loss rates and larger terminal velocities.
So the present calculations probably overestimate the mass loss rates. This may 
also be true for Vink \& Cassisi's results shown in 
Fig.~1. With the assumption $v_{\infty} = 5 v_{\rm esc}$ instead of $v_{\infty} = 
v_{\rm esc}$, their mass loss rates would be lower by a factor of the order of ten. 

The predicted mass loss rates shown in Fig.~1 clearly depend on the
surface gravity and the metallicity.  For $Z/ Z_{\odot} = 1$ coupled
winds can exist only if $\log g \la 5.5$. As the criterion for ion
decoupling is more restrictive than the one used in Sect.~2, this
requires total mass loss rates almost of the order $10^{-11} M_{\odot}
/ \rm yr$. If the metallicity is reduced by a factor of ten, then for
the considered effective temperatures at least for $\log g \ge 5.5$ no
solution of the momentum equation for a one component plasma exists at
all, because the radiative force is not sufficient.  This result
has been confirmed by calculations of Vink (priv. comm.).  For sdB
stars with $\log g = 5.5$ and $Z / Z_{\odot} = 1/10$ he could not find
any mass loss rate for which a global momentum balance is fulfilled.

\section{Discussion}
In the $T_{\rm eff} - \log g$ diagram of Fig.~2 for 
$25000 \mathrm{K} \leq T_{\rm eff} \leq 40000 \rm K$ and for various 
metallicities the lines are shown above which according to 
the mass loss calculations as described in Sect.~3 
chemically homogeneous winds may exist. 
\begin{figure}
\centering
\includegraphics[width=11cm,bb=132 75 625 425,clip=true]{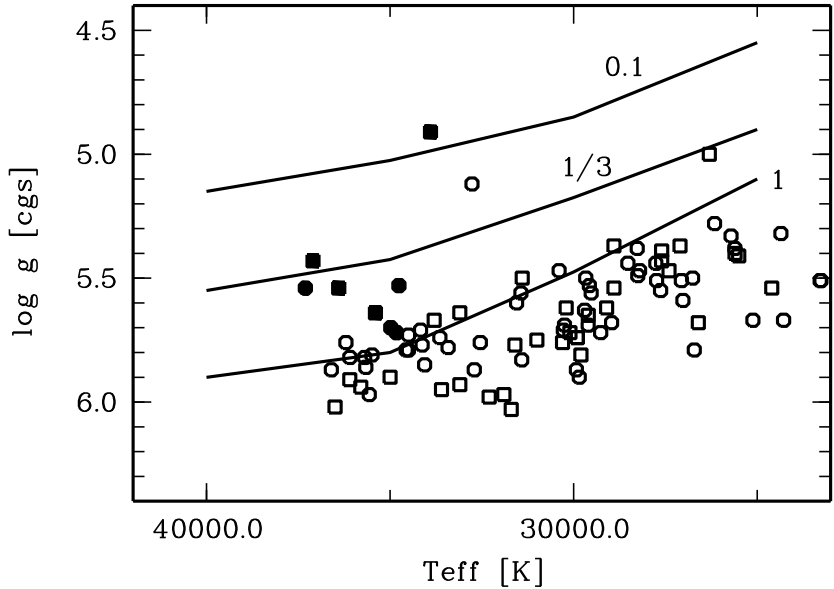}
\caption{Lines in the $T_{\rm eff} - \log g$ diagram above which 
chemically homogeneous winds may exist for $Z/Z_{\odot} = 0.1$, 
$1/3$ and $1$, respectively. Squares and circles represent the sdB stars 
analyzed 
by \citet{t12_max01} and \citet{t12_lisk05}, 
respectively. Filled symbols represent the sdB's with 
peculiar $\rm H \alpha$ line profiles, which may indicate the presence of 
a weak wind.}
\end{figure}
It can be seen that for the majority of sdB stars this is not possible
if the metallicity is solar or subsolar. They are below the line for
$Z / Z_{\odot} = 1$. The sdB stars introduced in the diagram have been
analyzed by \citet{t12_max01} and \citet{t12_lisk05}. According to the
assumptions of the present paper in those ones with peculiar $H
\alpha$ line profiles (represented by filled symbols) chemically
homogeneous winds may indeed exist if the metal abundances are not too
far below the solar value. The predicted mass loss rates agree with
the ones from the mass loss recipe of \citet{t12_vink02} and are of the
order $10^{-10}$ to $10^{-11} M_{\odot} / \rm yr$. However, as
explained in Sect.~3, this agreement does not exclude that the mass
loss rates are overestimated. So this result is still
questionable. The existence of winds may depend on the abundance of
one element only, which preferably contributes to the radiative
acceleration.

From the results it is clear that chemically homogeneous winds with mass loss 
rates $\dot M \la 10^{-12} M_{\odot} / \rm yr$ cannot exists. From arguments similar  
as in the paper of \citet{t12_ow2002}, this result can be obtained without the calculation 
of a wind model.
The mass loss rate of the metals alone must be at least of the order 
$10^{-14} M_{\odot} / \rm yr$. Otherwise hydrogen and helium cannot be 
accelerated throughout the wind. This result could be questioned 
only if a wind solution could be found for which the terminal velocity of
the metals is clearly lower than the surface escape velocity.
\citet{t12_krt00} suggested that the terminal velocity could indeed be 
lower than $v_{\rm esc}$ (by a factor of the order two only, however), 
if the wind switches to a ``shallow" solution with an abrupt lowering of 
the velocity gradient. However, \citet{t12_ow2002} and \citet{t12_krt02} argued that 
these solutions are unstable. 

If hydrogen and helium cannot be expelled from the star, then pure metallic winds 
may exist. As the outflowing metals in the stellar atmosphere not only have to 
overcome the gravitational force, but in addition the frictional force due to 
collisions with protons and helium particles, the metal abundances should be 
lower than predicted from equilibrium diffusion calculations (if concentration 
gradients are negligible). For these metals, for which the mass loss rate is 
sufficiently small, measured and predicted abundances should be in agreement. 
For metals with non-zero mass loss rate this scenario should lead to 
abundances varying with time as has been discussed by \citet{t12_sea96, t12_sea99} for 
iron group elements in the envelopes of HgMn stars.

If both hydrogen and helium are in hydrostatic equilibrium, then measured helium 
abundances should be approximately in agreement with the ones predicted from 
equilibrium diffusion calculations. The fact that in the various types of 
helium deficient chemically peculiar stars the measured ones are larger, seems 
to indicate that gravitational settling in general is less effective than expected 
in an undisturbed stellar atmosphere. Moreover the existence of two distinct 
sequences of sdB stars which are characterized by an offset in the helium 
abundance \citep{t12_edel03,t12_lisk05} can hardly be explained with one 
atmospheric effect alone. It may point to a dependence on the star's history.
Several scenarios of single star and binary evolution of the sdB and the hotter 
sdO stars are under discussion \citep{t12_str07}.
\acknowledgements{I gratefully acknowledge financial support by the 
organizers 
of the meeting and I thank I. Bues for carefully reading the manuscript.}


\begin{thebibliography}{}
\bibitem[Abbott(1982)]{t12_abb82}
Abbott, D. C. 1982, \apj, 259, 282
\bibitem[Adelman et al.(2006)]{t12_adel06}
Adelman, S. J., Caliskan, H., Gulliver, A. F., \& Teker, A. 
2006, \aap, 447, 685
\bibitem[Babel(1995)]{t12_bab95}
Babel, J. 1995, \aap, 301, 823
\bibitem[Babel(1996)]{t12_bab96}
Babel, J. 1996, \aap, 309, 867
\bibitem[Behara \& Jeffery(these proceedings)]{t12_nat08}
Behara, N. T., \& Jeffery, C. S. 2008, these proceedings
\bibitem[Bergeron et al.(1988)]{t12_ber88} Bergeron, P., Wesemael, F., 
Michaud, G., \& Fontaine, G. 1988, \apj, 332, 964
\bibitem[Blanchette et al.(2006)]{t12_blan06}
Blanchette, J. P., Chayer, P., Wesemael, F., et al. 2006, 
Baltic Astronomy, 15, 301
\bibitem[Burgers(1969)]{t12_bur69}
Burgers, J. M. 1969,  Flow equations for composite gases, 
(New York: Academic Press)
\bibitem[Castor, Abbott, \& Klein(1975, CAK)]{t12_ca1975}
Castor, J. I., Abbott, D. C., \& Klein, R. I. 1975, \apj, 195, 157 (CAK)
\bibitem[Charpinet et al.(1997)]{t12_char97} Charpinet, S., Fontaine, G., 
Brassard, P., et al. 1997, \apj, 483, L123
\bibitem[Chayer et al.(2004)]{t12_chay04}
Chayer, P., Fontaine, G., \& Fontaine, M. 2004, \apss, 291, 379
\bibitem[Chayer et al.(2006)]{t12_chay06} Chayer, P., Fontaine, F., 
Fontaine, G., Wesemael, F., \& Dupuis, J. 2006, Baltic Astronomy, 
15, 131
\bibitem[Edelmann et al.(2003)]{t12_edel03}
Edelmann, H., Heber, U., Hagen, H. J., et al. 2003, \aap, 400, 939
\bibitem[Edelmann et al.(2006)]{t12_edel06}
Edelmann, H., Heber, U., \& Napiwotzki, R. 2006, 
Baltic Astronomy, 15, 103
\bibitem[Fontaine et al.(2003)]{t12_fon03}
Fontaine, G., Brassard, P., Charpinet, S., et al. 
2003, \apj, 597, 518
\bibitem[Fontaine \& Chayer(1997)]{t12_fon97}
Fontaine, G., \& Chayer, P. 1997, in The Third Conference on Faint Blue 
Stars, eds. A. G. D. Philip, J. Liebert \& R. A. Saffer, 
(Schenectady: L. Davis Press), 169
\bibitem[Fontaine et al.(2006)]{t12_fon06}
Fontaine, G., Green, E. M., Chayer, P., et al. 
2006, Baltic Astronomy, 15, 211
\bibitem[Geier et al.(2008a)]{t12_gei08}
Geier, S., Heber, U., \& Napiwotzki, R. 2008a, \aap, in preparation
\bibitem[Geier et al.(2008b)]{t12_gei08b}
Geier, S., Heber, U., \& Napiwotzki, R. 2008b, these proceedings
\bibitem[Good et al.(2005)]{t12_good05}
Good, S. A., Barstow, M. A., Burleigh, M. R., et al.
2005, \mnras, 363, 183
\bibitem[Groth et al.(1985)]{t12_gro85}
Groth, H. G., Kudritzki, R. P., \& Heber, U. 1985, \aap, 152, 107
\bibitem[Hamann et al.(1981)]{t12_ham81}
Hamann, W. R., Gruschinske, J., Kudritzki, R. P., \& Simon, K. P. 
1981, \aap, 104, 249
\bibitem[Heber et al.(2003)]{t12_heb03}
Heber, U., Maxted, P. F. L., Marsh, T. R., Knigge, C., \& Drew, J. E.
2003, in ASP Conf.\ Ser.\ 288, Stellar Atmosphere Modeling, eds.\ I.~Hubeny, 
D.~Mihalas, \& K.~Werner,  (San Fransicsco: ASP), 251
\bibitem[Heber et al.(2000)]{t12_heb00}
Heber, U., Reid, I. N., \& Werner, K. 2000, \aap, 363, 198
\bibitem[Jomaron et al.(1999)]{t12_jom99}
Jomaron, C. M., Dworetsky, M. M., \& Allen, C. 1999, \mnras, 303, 555
\bibitem[Krti\v cka(2006)]{t12_krt06}
Krti\v cka, J. 2006, \mnras, 367, 1282
\bibitem[Krti\v cka \& Kub\'at(2000)]{t12_krt00}
Krti\v cka, J., \& Kub\'at, J. 2000, \aap, 359, 983
\bibitem[Krti\v cka \& Kub\'at(2002)]{t12_krt02}
Krti\v cka, J., \& Kub\'at, J. 2002, \aap, 388, 531
\bibitem[Krti\v cka et al.(2003)]{t12_krt03}
Krti\v cka, J., Owocki, S. P., Kub\'at, J., Galloway, R. K., 
\& Brown, J. C. 2003, \aap, 402, 713
\bibitem[Kudritzki(2002)]{t12_ku2002}
Kudritzki, R. P. 2002, \apj, 577, 389
\bibitem[Lamers \& Cassinelli(1999)]{t12_la1999}
Lamers, H. J. G. L. M., \& Cassinelli, J. P. 1999, Introduction to Stellar 
Winds, (Cambridge: Cambridge University Press)
\bibitem[Lisker et al.(2005)]{t12_lisk05}
Lisker, T., Heber, U., Napiwotzki, R., et al. 2005, \aap, 430, 223
\bibitem[Maxted et al.(2001)]{t12_max01}
Maxted, P. F., Heber, U., Marsh, T. R., \& North, R. C.
2001, \mnras, 326, 1391
\bibitem[Michaud et al.(1989)]{t12_mic89}
Michaud, G., Bergeron, P., Heber, U., \& Wesemael, F. 
1989, \apj, 338, 417
\bibitem[Michaud et al.(1979)]{t12_mic79}
Michaud, G., Montmerle, T., Cox, A. N., et al. 1979, \apj, 234, 206
\bibitem[Napiwotzki(1999)]{t12_napi99} 
Napiwotzki, R. 1999, \aap, 305, 301
\bibitem[Ohl et al.(2000)]{t12_ohl00} Ohl, R. G., Chayer, P., \& Moos, H. W. 2000, 
         \apj, 538, L95
\bibitem[O'Toole et al.(2005)]{t12_otol05}
O'Toole, S. J., Jordan, S., Friedrich, S., \& Heber, U. 
2005, \aap, 437, 227
\bibitem[O'Toole \& Heber(2006)]{t12_otol06}
O'Toole, S. J., \& Heber, U. 2006, \aap, 452, 579
\bibitem[Owocki \& Puls(2002)]{t12_ow2002}
Owocki, S. P., \& Puls, J. 2002, \apj, 568, 965	
\bibitem[Proffitt et al.(1999)]{t12_prof99}
Proffitt, C. R., Brage, T., Leckrone, D., et al. 1999, \apj, 512, 942
\bibitem[Puls et al.(2000)]{t12_puls00}
Puls, J., Springmann, U., \& Lennon, M. 2000, \aaps, 141, 23
\bibitem[Rauch(1993)]{t12_rau93}
Rauch, T. 1993, \aap, 276, 171
\bibitem[Roby et al.(1999)]{t12_roby99}
Roby, S. W., Leckrone, D. S., \& Adelman, S. J. 1999, \apj, 524, 974
\bibitem[Schuh et al.(2005)]{t12_son05}
Schuh, S. L., Barstow, M. A., \& Dreizler, S. 2005, in ASP Conf. Ser.
334, eds.\ D. Koester \& S. Moehler, (San Francsico: ASP), 237
\bibitem[Schuh et al.(2002)]{t12_son02}
Schuh, S. L., Dreizler, S., \& Wolff, B. 2002, \aap, 382, 164
\bibitem[Seaton(1996)]{t12_sea96}
Seaton, M. J. 1996, \apss, 237, 107
\bibitem[Seaton(1999)]{t12_sea99}
Seaton, M. J. 1999, \mnras, 307, 1008
\bibitem[Smith(1996)]{t12_smith96}
Smith, K. C. 1996, \apss, 237, 77
\bibitem[Springmann \& Pauldrach(1992)]{t12_spr92}
Springmann, U. W. E., \& Pauldrach, A. W. A. 1992, \aap, 262, 515
\bibitem[Stroeer et al.(2007)]{t12_str07}
Stroeer, A., Heber, U., Lisker, T., et al. 2007, \aap, 462, 269
\bibitem[Turcotte(2003)]{t12_tur03}
Turcotte, S. 2003, ASPC, 305, 199
\bibitem[Unglaub(A\&A, submitted)]{t12_ku08}
Unglaub, K. 2008, \aap, submitted
\bibitem[Unglaub \& Bues(1998)]{t12_ub98}
Unglaub, K., \& Bues, I. 1998, \aap, 338, 75
\bibitem[Unglaub \& Bues(2001)]{t12_ub01}
Unglaub, K., \& Bues, I. 2001, \aap, 374, 570
\bibitem[Vauclair et al.(1978)]{t12_vauc78}
Vauclair, G., Vauclair, S., \& Michaud, G. 1978, \apj, 223, 920
\bibitem[Vauclair et al.(1979)]{t12_vauc79}
Vauclair, G., Vauclair, S., \& Greenstein, J. L. 1979, \aap, 80, 79
\bibitem[Vennes et al.(1988)]{t12_ven88}
Vennes, S., Pelletier, C., Fontaine, G., \& Wesemael, F. 
1988, \apj, 331, 876
\bibitem[Vink(2004)]{t12_vink04}
Vink, J. S. 2004, \apss, 291, 239
\bibitem[Vink \& Cassisi(2002)]{t12_vink02}
Vink, J. S., \& Cassisi, S. 2002, \aap, 392, 553
\bibitem[Vink et al.(2001)]{t12_vink01}
Vink, J., de Koter, A., \& Lamers, H. J. G. L. M. 
2001, \aap, 369, 574
\bibitem[Zavala et al.(2007)]{t12_zav07}
Zavala, R. T., Adelman, S. J., \& Hummel, C. A. 2007, \apj, 655, 1046
\end{thebibliography}
\end{document}